\documentclass[fleqn,usenatbib]{mnras}

\usepackage{newtxtext,newtxmath}

\usepackage[T1]{fontenc}

\DeclareRobustCommand{\VAN}[3]{#2}
\let\VANthebibliography\thebibliography
\def\thebibliography{\DeclareRobustCommand{\VAN}[3]{##3}\VANthebibliography}

\usepackage{graphicx}	
\usepackage{amsmath}	

\title[JWST 7- 10- 15- micron source counts]{
Galaxy source counts at 7.7\, $\mu$m, 10\, $\mu$m and 15\,$\mu$m with the James Webb Space Telescope}

\author[Ling et al. 2022]{
Chih-Teng Ling$^{1}$,
Seong Jin Kim$^{1}$,
Cossas K.-W. Wu$^{1,2}$,
Tomotsugu Goto$^{1,2}$, 
Ece Kilerci$^{3}$,
\newauthor
Tetsuya Hashimoto$^{4}$,
Yu-Wei Lin$^{1,2}$,
Po-Ya Wang$^{1}$,
Simon C.-C. Ho$^{1}$, and 
Tiger Yu-Yang Hsiao$^{1}$
\\
$^{1}$Institute of Astronomy, National Tsing Hua University, 101, Section 2. Kuang-Fu Road, Hsinchu, 30013, Taiwan (R.O.C.)\\
$^{2}$Department of Physics, National Tsing Hua University, 101, Section 2. Kuang-Fu Road, Hsinchu, 30013, Taiwan (R.O.C.)\\
$^{3}$Sabanc{\i} University, Faculty of Engineering and Natural Sciences, 34956, Istanbul, Turkey\\
$^{4}$Department of Physics, National Chung Hsing University, 145, Xingda Road, Taichung, 40227, Taiwan (R.O.C.)
}

\date{Accepted XXX. Received YYY; in original form ZZZ}

\pubyear{2022}

\begin{document}
\label{firstpage}
\pagerange{\pageref{firstpage}--\pageref{lastpage}}
\maketitle

\begin{abstract}
We present mid-infrared galaxy number counts based on the Early Release Observations obtained by the James Webb Space Telescope (\textit{JWST}) at 7.7-, 10- and 15-$\mu$m (F770W, F1000W and F1500W, respectively) bands of the Mid-Infrared Instrument (MIRI). Due to the superior sensitivity of \textit{JWST}, the 80 percent completeness limits reach 0.32, 0.79 and 2.0 $\mu$Jy in F770W, F1000W and F1500W filters, respectively, i.e., $\sim$100 times deeper than previous space infrared telescopes such as \textit{Spitzer} or \textit{AKARI}. The number counts reach much deeper than the broad bump around $0.05\sim0.5$ mJy due to polycyclic aromatic hydrocarbon (PAH) emissions. 
An extrapolation towards fainter flux from the evolutionary models in the literature agrees amazingly well with the new data, where the extrapolated faint-end of infrared luminosity functions combined with the cosmic star-formation history to higher redshifts can reproduce the deeper number counts by \textit{JWST}. Our understanding of the faint infrared sources has been confirmed by the observed data due to the superb sensitivity of \textit{JWST}.
\end{abstract}




\section{Introduction}
The James Webb Space Telescope \citep[\textit{JWST},][]{Gardner2006, Kalirai2018} is a new generation infrared (IR) telescope launched in December 2021, and has just opened up a new window to observe the faintest IR source populations that have not been observed yet. The infrared space telescopes in the pre-JWST era, namely the Infrared Astronomical satellite \citep[\textit{IRAS},][]{Neugebauer1984}, \textit{ISO} \citep{Kessler1996}, \textit{AKARI} \citep{Murakami2007}, \textit{Spitzer} \citep{Werner2004}, and \textit{Herschel} \citep{Pilbratt2010} have showed the evolution of the luminosity functions (LFs) of infrared galaxy populations in luminosity and density  \citep[e.g.,][]{Saunders1990,Rowan-Robinson1997,Elbaz2002,Caputi2007,Goto2010, Gruppioni2010, Gruppioni2011}.
The comparison of the predictions of galaxy formation models with the observed number density of infrared galaxies is a very useful method to understand their formation and evolution history. Therefore, source counts obtained from a deep observation are crucial to understanding the evolution of these galaxies.

The mid-IR (MIR) selected galaxies have been classified roughly into three types according to the predominant emission component contributing to the MIR range of the spectral energy distribution: (i) star-forming (SF) galaxies including starbursts with various MIR (e.g., 8 $\mu$m or 12 $\mu$m) luminosities between $10^{8}-10^{11} L_{\odot}$, %
(ii) active galaxies including type 1 and type 2 active galactic nuclei (AGNs), and (iii) composites, the mixture of the emissions from the star-forming regions and AGN activity (i.e., hot dusty torus). 
Mid-IR number counts of galaxies have been obtained in AKARI bands \citep[e.g.,][]{Wada2008, Pearson2010, Pearson2014, Takagi2012}, ISO bands \citep[e.g.,][]{Pearson2005}, Spitzer bands \citep[e.g.,][]{Pearson2005}. Although the lower end of the mid-IR number counts, luminosity functions (LFs), and evolution models have not been fully investigated in the pre-JWST era, they have been predicted in several studies \citep[e.g.,][]{Gruppioni2011, Cowley2018, Shen2022}. 

The Mid-Infrared Instrument \citep[MIRI,][]{Rieke2015b} of \textit{JWST} performs imaging \citep[][]{Bouchet2015} with 9 bands covering the 5-28 $\mu$m wavelength range. 
In this work, our goal is to obtain the first galaxy number counts in 7.7 $\mu$m, 10.0 $\mu$m, and 15.0 $\mu$m from the MIR data of \textit{JWST}.  
We aim to characterise the evolution of the MIR LF and 
luminosity density (LD) up to a new sensitivity limit that is just available with \textit{JWST}. 
This paper is organised as the following: 
In \S \ref{S:data} we present the \textit{JWST} data, source extraction and the completeness of our source detection.
In \S \ref{S:res}, we compare the observed source counts with the model predictions from the literature and discuss our results.
Our conclusion is given in \S \ref{S:conc}.
We follow \citet{Cowley2018} whose model we compare with, adopting the {\it Planck15} cosmology \citep{Planck2016}, i.e., $\Lambda$ cold dark matter cosmology with ($\Omega_{m}$,$\Omega_{\Lambda}$,$\Omega_{b}$,$h$)=(0.307, 0.693, 0.0486, 0.677), unless otherwise mentioned. 

\section{Data Analysis}\label{S:data}
\subsection{Source extraction}
We use the publicly available \textit{JWST} Early Release Observations (ERO) image of Stephan's quintet\footnote[1]{\url{https://www.stsci.edu/jwst/science-execution/approved-programs/webb-first-image-observations}} observed in F770W, F1000W and F1500W broadband filters centred at 7.7 $\mu$m, 10.0 $\mu$m and 15.0 $\mu$m wavelengths, respectively. 
With \textit{JWST} MIRI, Stephan's quintet is imaged separately in two components, Hickson Compact Group 92 and a foreground galaxy NGC 7320.
The IDs of the two images are $\text{jw02732-o002\_t001}$ and $\text{jw02732-o006\_t001}$, respectively. Hereafter we use '002' and '006' for $\text{jw02732-o002\_t001}$ and $\text{jw02732-o006\_t001}$, respectively.

\begin{figure}
    \centering
    \includegraphics[width=\columnwidth]{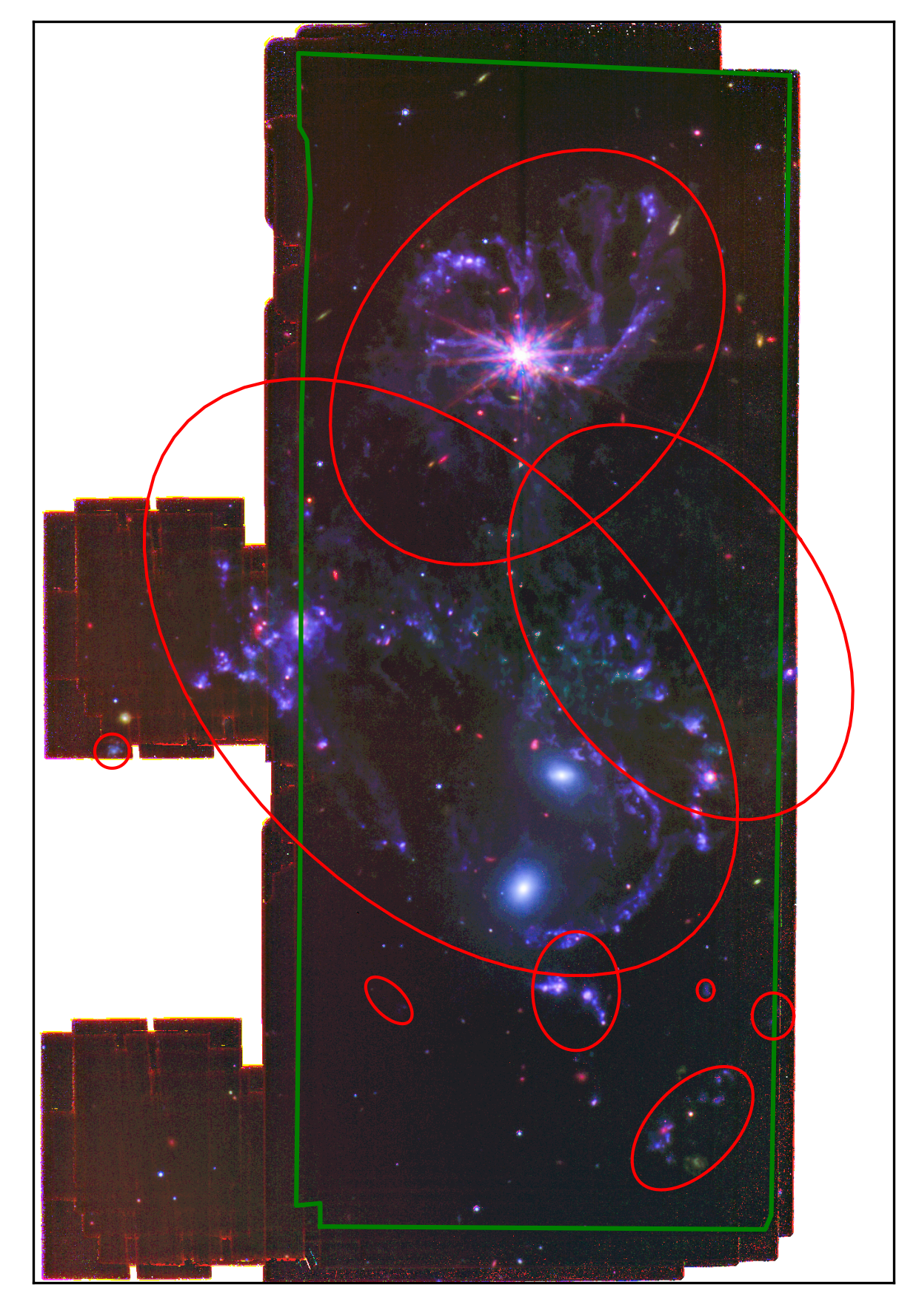}
    \includegraphics[width=\columnwidth]{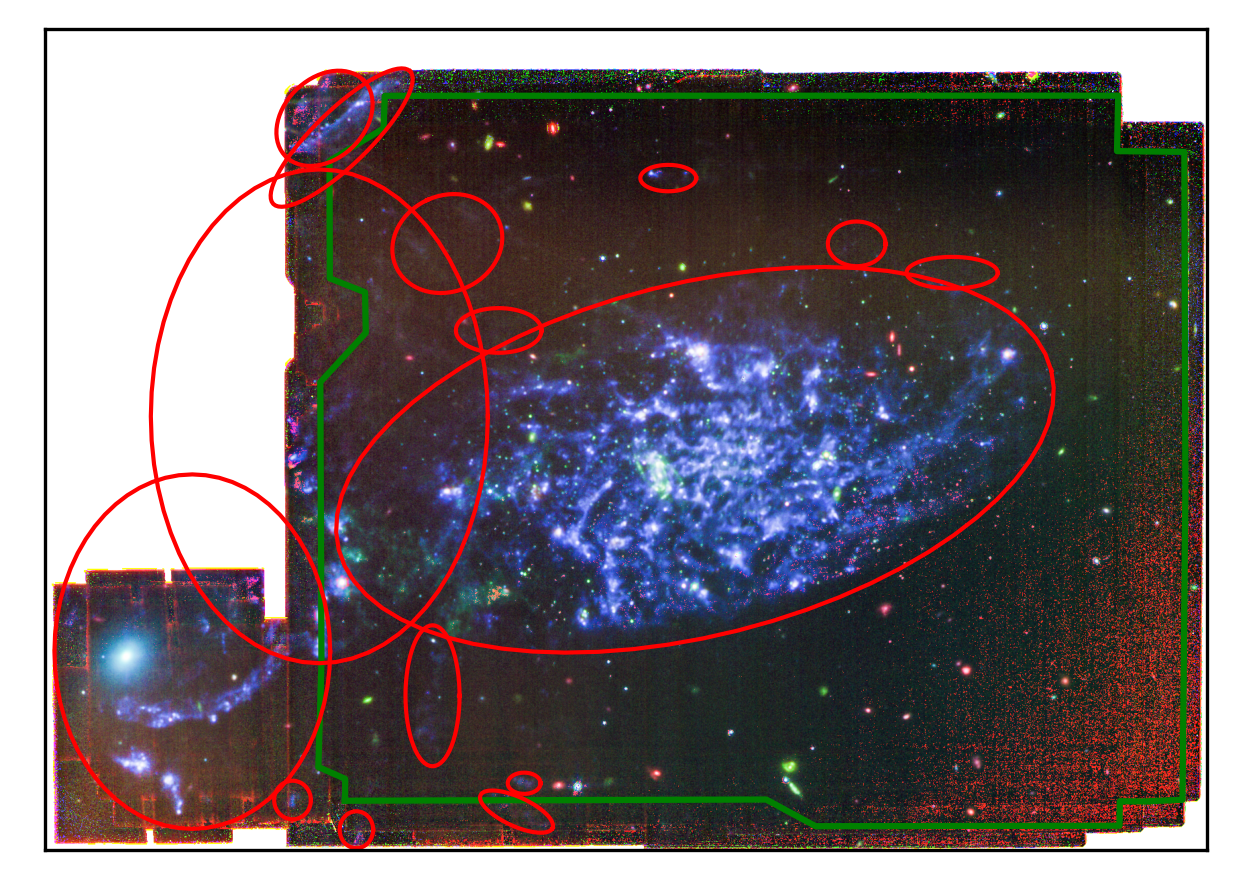}
    \caption{
        The mask on the false-colour images of Stephan's quintet with red (F1500W), green (F1000W) and blue (F770W). 
        Upper: image 002; lower: image 006. 
        Regions within red ellipses and outside green boxes are masked and not used in our analysis. 
        The same mask is applied for all 3 filters.
    }
    \label{fig:1}
\end{figure}

Since we aim to detect sources to the faintest limit observed with these filters, we perform our own source extraction as described in the following paragraph rather than using the public source catalogue. 
First, we mask Stephan's quintet in the centre of the images and foreground extended objects to define the field regions. The masked parts of the images are shown in Figure~\ref{fig:1}, where regions within red ellipses and outside green boxes are masked and discarded.
After removing foreground objects, we have 6924 arcsec$^2$ and 9764 arcsec$^2$ sky coverage for image 002 and image 006, respectively.
We utilise \textsc{Source-Extractor} V2.19.5 \citep{Bertin1996} to extract sources in these fields. 
Following the \textit{JWST} public catalogue, we measure source fluxes from apertures that include 70\% energy of the simulated \textit{JWST} MIRI point spread function \citep[PSF,][]{WebbPSF2014} for better statistics of the source counts.
Parameters that are specified in \textsc{Source-Extractor} are listed in Table \ref{table:1}. 
DETECT\_THRESH in Table \ref{table:1} stands for how many times the pixel value above the root mean square (RMS) of the background would be considered as a detected source in \textsc{Source-Extractor}. A value of 1.75 is applied here for every filter to detect sources that are not picked up by the official catalogue. The background RMS estimation is determined by BACK\_SIZE and BACKPHOTO\_THICK in Table \ref{table:1}. The former indicates the size of the neighbourhood of the source to be used in background calculations and the latter affects the width of the annulus of background estimation in pixel unit.
The pixel scale is 0.1109 arcsec in all filters.
Other unmentioned parameters remain the default value of \textsc{Source-Extractor}.
Please refer the manual \citep{Holwerda2005} and the document\footnote[2]{\url{https://sextractor.readthedocs.io/en/latest/index.html}} of \textsc{Source-Extractor} for the detail of other parameters.

\begin{figure}
    \centering
    \includegraphics[width=\columnwidth]{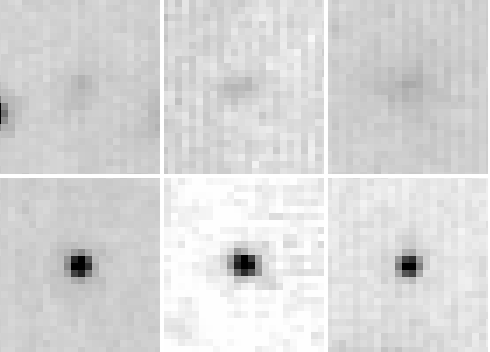}
    \caption{Examples of detected sources in field 002 at the 7.7-$\mu$m band (F770W). The fluxes of objects are $\sim0.40~\mu$Jy  in the first row and $\sim4.0~\mu$Jy in the second row. All images have a field-of-view of 2$\times$2 arcsec$^{2}$.}
    \label{002f770source}
\end{figure}

\begin{figure}
    \centering
    \includegraphics[width=\columnwidth]{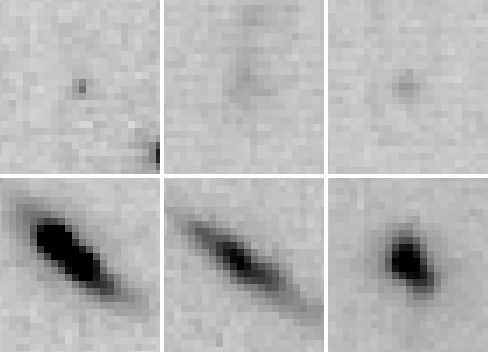}
    \caption{Examples of detected sources in field 002 at the 10-$\mu$m band (F1000W). The fluxes of objects are $\sim0.79~\mu$Jy  in the first row and $\sim7.9~\mu$Jy in the second row. All images have a field-of-view of 2$\times$2 arcsec$^{2}$.}
    \label{002f1000source}
\end{figure}

\begin{figure}
    \centering
    \includegraphics[width=\columnwidth]{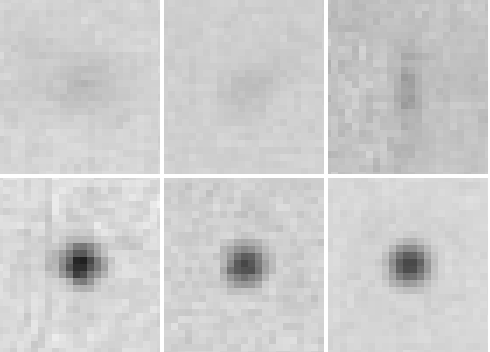}
    \caption{Examples of detected sources in field 002 at the 15-$\mu$m band (F1500W). The fluxes of objects are $\sim2.0~\mu$Jy  in the first row and $\sim20~\mu$Jy in the second row. All images have a field-of-view of 2$\times$2 arcsec$^{2}$.}
    \label{002f1500source}
\end{figure}

Examples of extracted sources in each filter from image 002 are presented in Figures ~\ref{002f770source}, ~\ref{002f1000source} and ~\ref{002f1500source}. 
These objects are randomly selected by their fluxes.
The fluxes of objects in the first row are about the 80\% completeness limit of each filter (as described in Section \ref{completeness}), while the fluxes of objects in the second row are around ten times higher than those in the first.

\begin{table}
    \centering
    \begin{tabular}{ccp{0.54\columnwidth}}
        \hline
        Parameter Name && Common to the all filters \\
        \hline
        DETECT$\_$THRESH && 1.75 \\
        DEBLEND$\_$NTHRESH && 48 \\
        DEBLEND$\_$MINCONT && 0.0008 \\
        BACK$\_$SIZE && 3 \\
        BACK$\_$FILTERSIZE && 3 \\
        BACKPHOTO$\_$TYPE && LOCAL \\
        BACKPHOTO$\_$THICK && 12 \\
        FILTER$\_$NAME && gauss$\_$2.5$\_$5$\times$5.conv
    \end{tabular}

    \begin{tabular}{cp{0.15\columnwidth}p{0.15\columnwidth}p{0.15\columnwidth}}
        \hline
         & F770W & F1000W & F1500W \\
        \hline 
        PHOT$\_$APERTURES [pix] & 5.78 & 7.62 & 11.49 \\
        \hline
        
    \end{tabular}
    \caption{Parameters specified in \textsc{Source-Extractor}.}
    \label{table:1}
\end{table}

\subsection{Completeness}\label{completeness}
The completeness of the source detection is not provided in the public \textit{JWST} source catalogue.
Therefore, we estimate the completeness as a function of flux density based on our source extraction.
The derived source count is corrected based on completeness. 
To estimate the completeness of our extracted source catalogue, we adopt the artificial source method described in \cite{Takagi2012}.
We randomly distribute 20 artificial point sources in each image, avoiding placement on the image edges or on the masks.
The sources are generated from simulated \textit{JWST} MIRI PSF, where we assume a flat spectrum. 
For each filter in each image, we executed the \textsc{Source-Extractor} in the same way as we performed the source detection and quantify the fractions of artificial sources recovered. 
This process is repeated 400 times with varying flux ranging from 10 nJy to 0.1 mJy in each filter and each image, with a total of 8000 artificial sources implanted.
This number is comparable to \cite{Takagi2012}, considering the difference in the sky coverage.

\begin{figure}
    \includegraphics[width=\columnwidth]{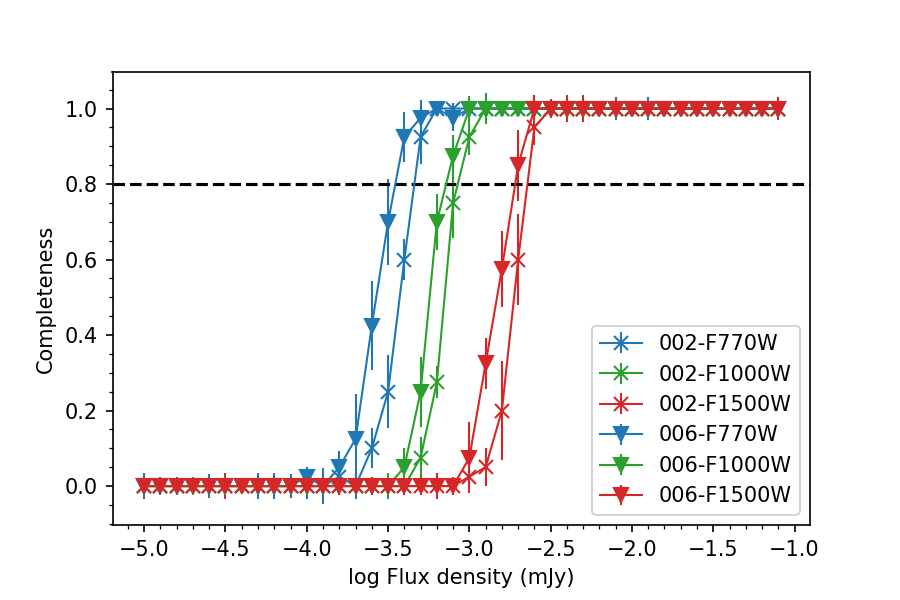}
    \caption{Source completeness of the field regions in Stephan's quintet images with a bin width of $\Delta\log (f_{\nu}/{\rm Jy})=0.1$ dex. Asterisk and triangle markers indicate the images, 002 and 006, respectively. The different filters are plotted in blue (F770W), green (F1000W), and red (F1500W). The black dashed line shows our criterion of the 80\% completeness. }
    \label{fig:complete}
\end{figure}

The obtained completeness is presented in Figure \ref{fig:complete}, wherein the 006 field is slightly deeper than the 002 field in all three filters. In the 002 (006) fields, the 80 percent completeness limit reaches 0.50 (0.32), 0.79 (0.79), and 2.51 (2.00) $\mu$Jy in F770W, F1000W and F1500W filters, respectively.
These are slightly better than the sensitivity values in the MIRI instrument handbook\footnote[3]{\url{http://web.physics.ucsb.edu/~cmartin/data/4clm/MIRI._Cycle1.pdf}}, where the point source detection limits (SN $=10$) are 
0.58, 1.2, and 3.4 $\mu$Jy, respectively, when scaled to the 2664 sec exposure time for each filter.
In the following source count analysis described in Section \ref{S:res}, the number count of each flux bin is weighted by its completeness. However, we only use sources up to 80 percent of completeness limits.


\begin{figure}
  \begin{center}
  \includegraphics[width=\columnwidth]{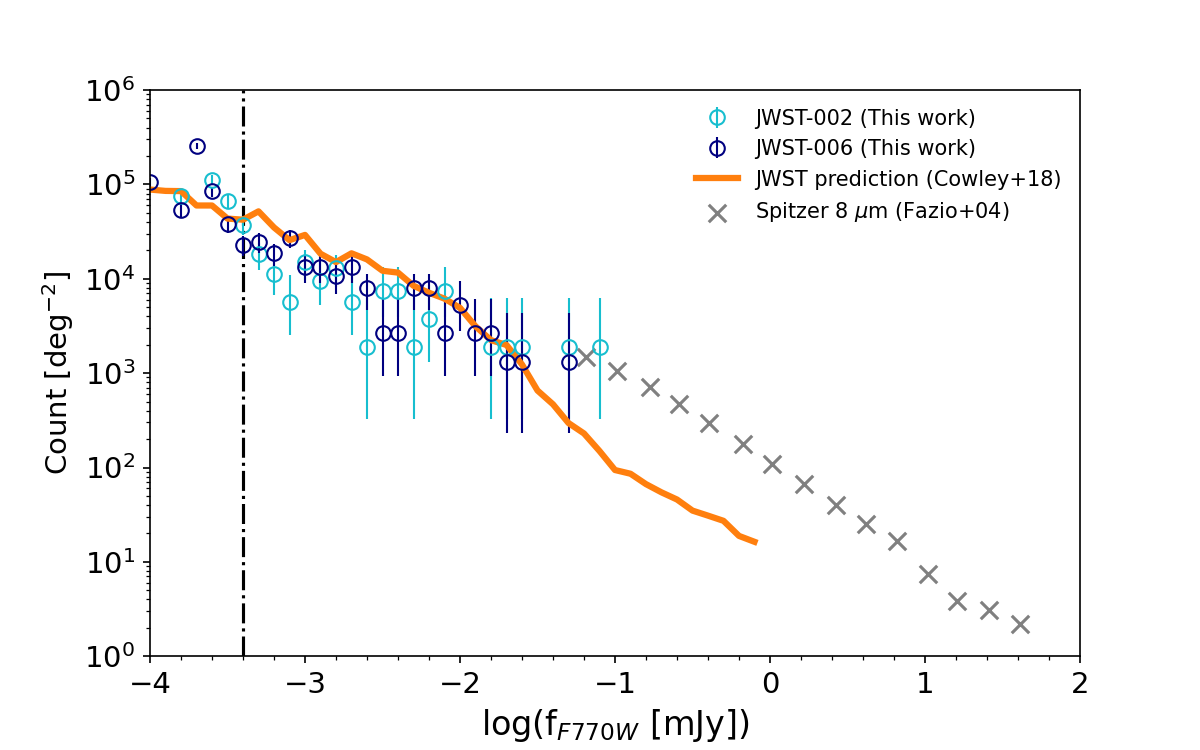}
  \includegraphics[width=\columnwidth]{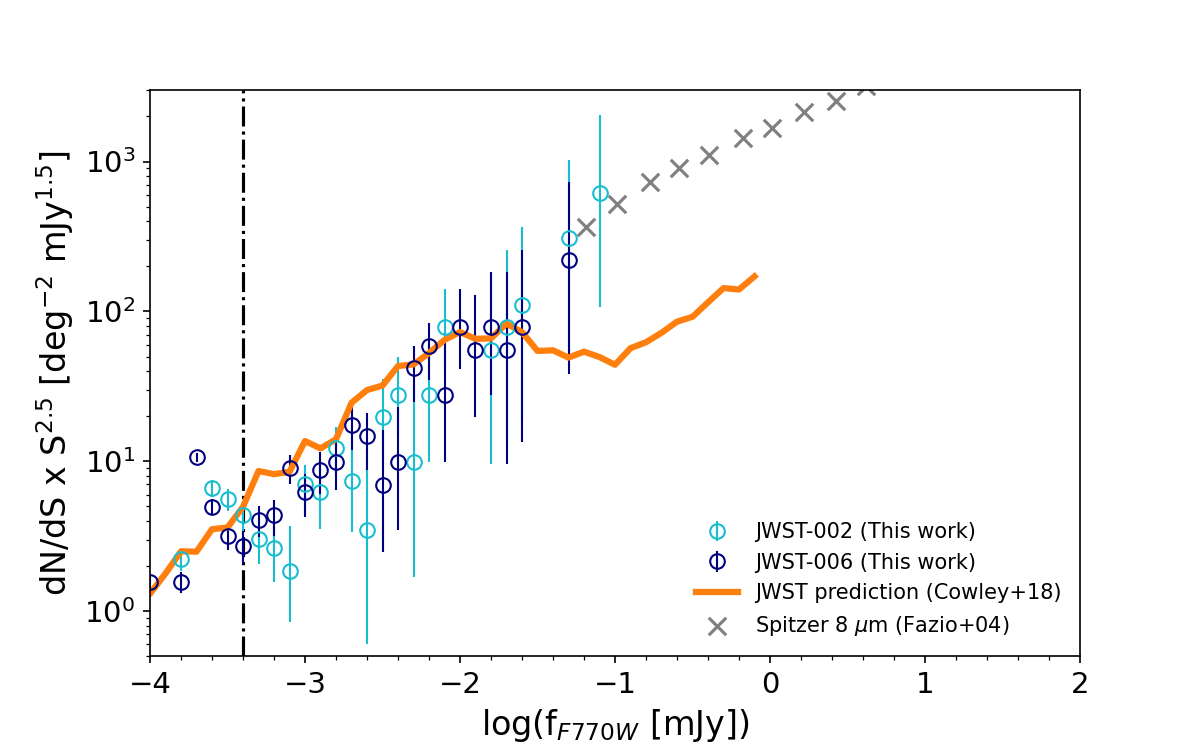}
    \caption{Source counts and models in the 7-$\mu$m band (F0700W). 
    The top panel shows the number of sources per deg$^{2}$ in each flux bin of $\Delta\log (f_{\nu}/{\rm Jy})=0.1$ dex.  
    Open circles indicate the source counts using the \textit{JWST} data in this work. 
    The orange line shows the prediction (differential count) from \citet{Cowley2018}. 
    We also compare with the \textit{Spitzer} 8 $\mu$m source counts for the Bo\"{o}tes field ($\times$ symbol) from \citet{Fazio2004}, where stars dominate the bright end. 
    The vertical (dot-dashed) line indicates the \textit{JWST} 80$\%$ completeness level of the source detection (002 field) derived in this work. 
    The bottom panel shows the differential source counts normalized to the Euclidean space.  } 
    \label{SC_07}
   \end{center}    
\end{figure}

\begin{figure}
  \begin{center}
  \includegraphics[width=\columnwidth]{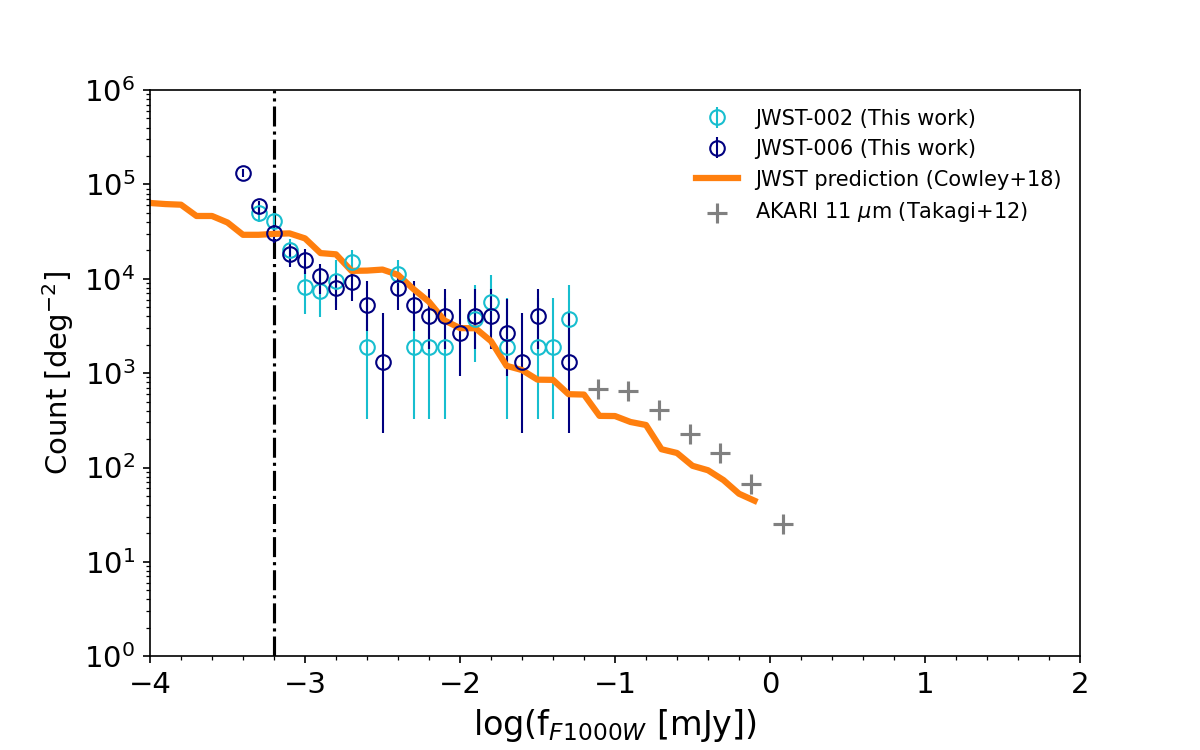}
  \includegraphics[width=\columnwidth]{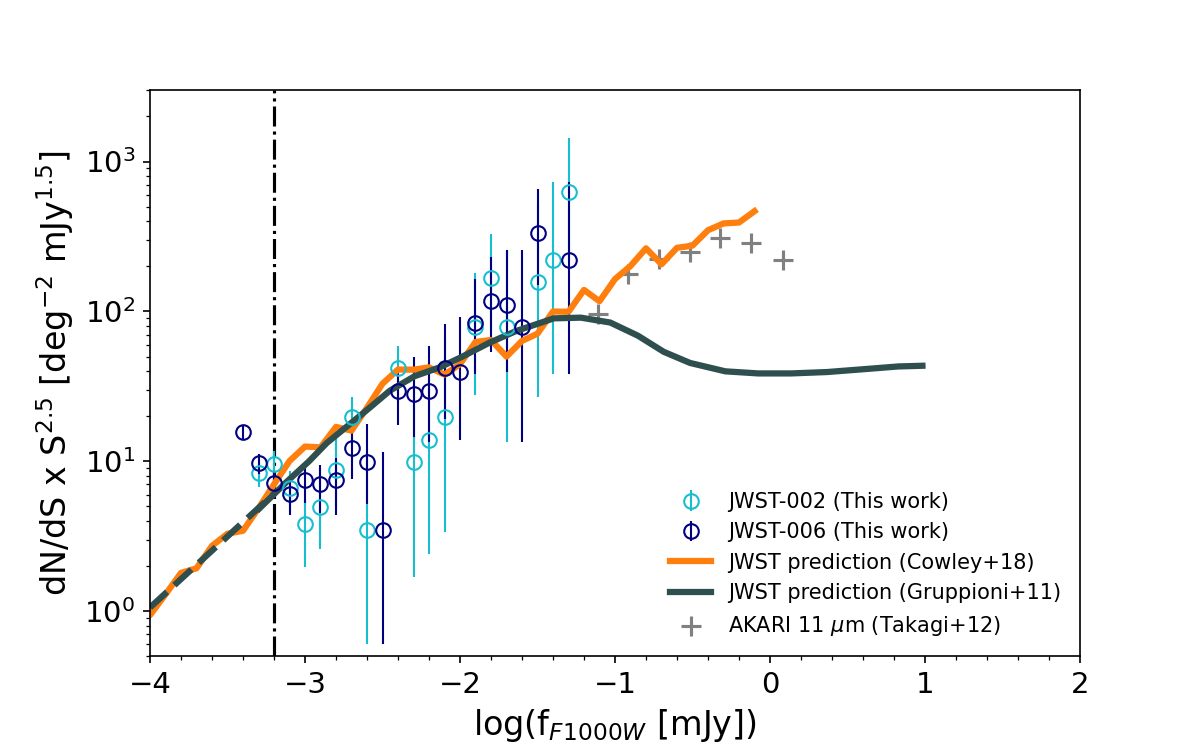}
    \caption{
    Same as Figure \ref{SC_07}, but for 10-$\mu$m band (F1000W).
    The dark grey curve from \citet{Gruppioni2011} is consistent with the observed number counts (open circles), where we linearly extrapolate the model prediction to the low-flux end (shown as a dark grey dashed line). 
    We also compare with the 11$\mu$m (S11 band of \textit{AKARI}) source counts ($+$ symbol) from \citet{Takagi2012}.   The vertical (dot-dashed) line indicates the \textit{JWST} 80$\%$ completeness level of the source detection (002 field) derived in this work. 
    }
    \label{SC_10}
 \end{center}    
\end{figure}

\begin{figure}
  \begin{center}
  \includegraphics[width=\columnwidth]{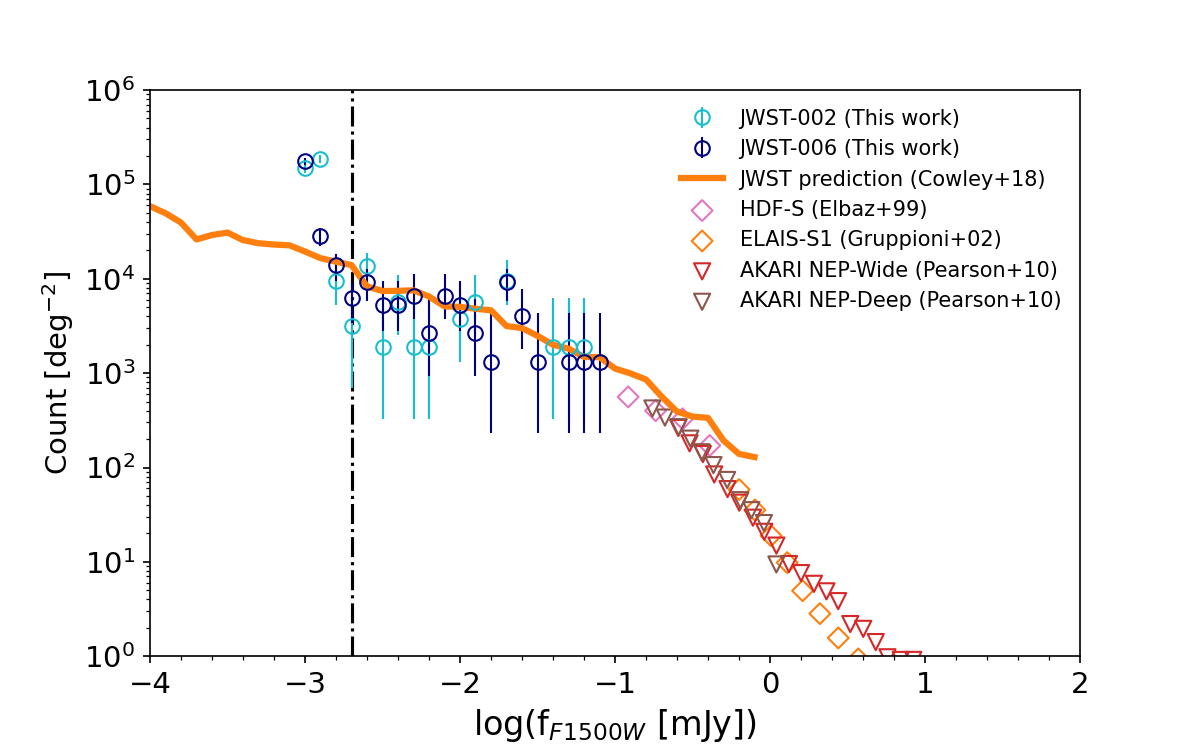}
  \includegraphics[width=\columnwidth]{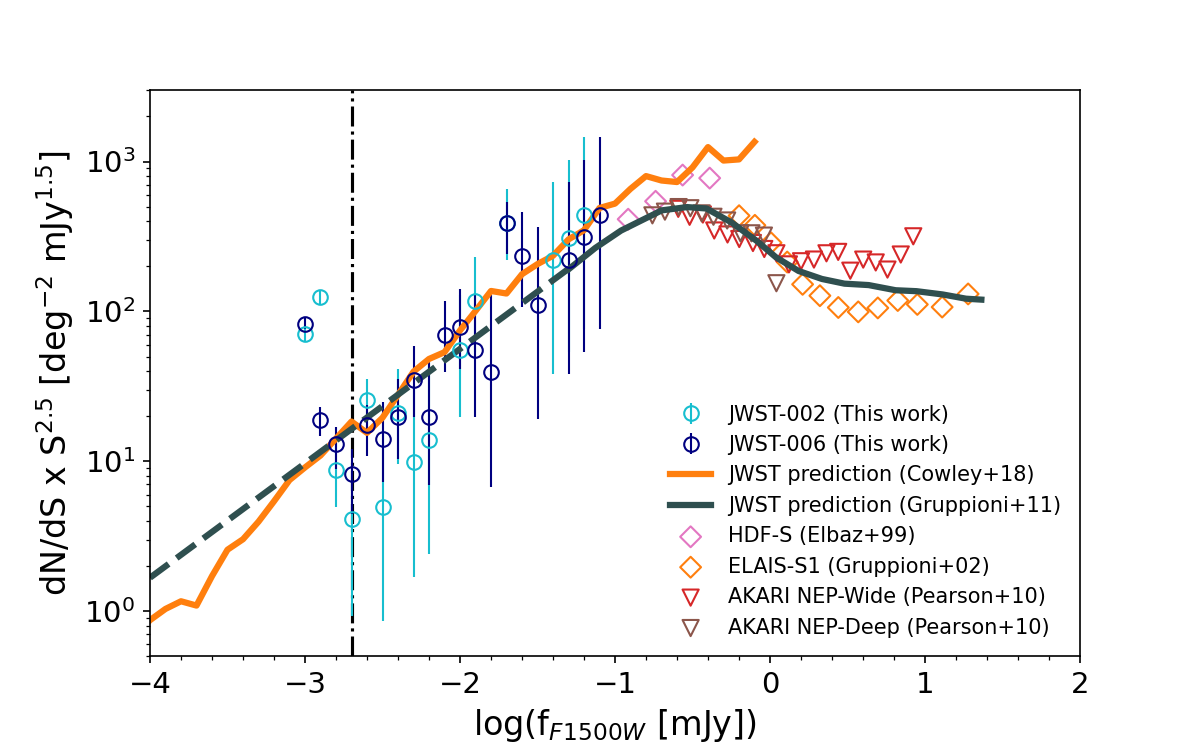}
    \caption{
    Same as Figure \ref{SC_07}, but for 15-$\mu$m band (F1500W).
    The dark grey curve from \citet{Gruppioni2011} shows a reasonable agreement with our observed number counts (open circles), where we linearly extrapolate the model prediction to the low-flux end (shown as a dark grey dashed line). 
    The previous source counts from \citet[][pink diamonds]{Elbaz1999}, \citet[][orange diamonds]{Gruppioni2002} and \citet[][inverted triangles]{Pearson2010} are also compared.  The vertical (dot-dashed) line indicates the \textit{JWST} 80$\%$ completeness level of the source detection (002 field) derived in this work. 
   }
    \label{SC_15}
  \end{center}
\end{figure}

\section{Results and discussion}\label{S:res}
We present the derived number counts in light/dark blue hollow circles (002/006 fields) in Figures \ref{SC_07}-\ref{SC_15}. 
The error bars of our number counts are obtained based on \citet{Gehrels1986}. 

To compare with the observed source counts, we overplot source count models for MIRI's F1000W and F1500W filter bands, taken from Figures 14 and 5 of \citet{Gruppioni2011}, respectively. They employed local luminosity functions (LLFs) and backward evolution for galaxies and AGNs. 
The 10- and 15-$\mu$m LLFs parameters are determined for five representative populations of IR galaxies, including spiral galaxies, starburst galaxies, low-luminosity AGNs, type 1 AGNs and type 2 AGNs, based on the previous \textit{Spitzer} and \textit{Hershel} data. The luminosity and density evolution follow their equations 2 and 3, with the model parameters presented in their Table 2.  
We also overplot model predictions from \citet{Cowley2018}.
They started from the GALFORM model \citep{Cole2000}, embedded within a dark matter only simulation (800 Mpc$^{3}$) with a halo mass resolution of M$_{\rm halo} > 2\times$10$^{9}$ h$^{-1}$ M$_{\odot}$ provided by \cite{Baugh2019MNRAS.483.4922B}. 
To compute SEDs, they used the spectrophotometric radiative transfer code \textsc{GRASIL} \citep{Silva1998}.

Immediately noticeable in our results is the depths of \textit{JWST}. While the brightest data points from \textit{JWST} (this work) are around 0.1 mJy, it is already comparable to the deepest flux densities of the overplotted previous works from \textit{Spitzer} \citep{Fazio2004} and \textit{AKARI} \citep{Takagi2012}.
The \textit{JWST} data continue to $\sim~\mu$Jy levels. 
This has been shown in Figures \ref{002f770source}-\ref{002f1500source}, where the sources near (and ten times brighter) the \textit{JWST} 80\% completeness limit of each filter are all reliably detected in those post-stamp images. 
The source counts obtained in this work are well connected to the result from previous observations, \citep[i.e.,][]{Fazio2004, Takagi2012, Elbaz1999, Gruppioni2002, Pearson2010}, and extends the source counts to about two orders below than the literature.

 In Figures \ref{SC_10} and \ref{SC_15}, the derived number counts agree well with the model prediction for \textit{JWST} in the dark grey line \citep{Gruppioni2011}, which were based on the infrared data from the previous generation satellites \citep[\textit{Spitzer}/\textit{AKARI};][]{Gruppioni2010}. 
 Furthermore, our observational results also agree well with the predictions from the dark matter simulation with an embedded semi-analytical calculation \citep{Cowley2018} in all Figures \ref{SC_07}-\ref{SC_15}.
 The agreements show that our understanding of the faint population of infrared galaxies based on the previous data was correct. It is a great achievement of \textit{JWST} to confirm models by directly detecting faint population of infrared galaxies undetectable with previous telescopes ($<$0.1 mJy).

We find that number counts agree well with previous empirical models. However, the models for the number counts are derived from a number of parameters such as luminosity evolution, density evolution, and the faint-end of LFs. These parameters degenerate in the number counts. The next important step is to disentangle these parameters by using photometric redshifts, and SED fitting techniques. The evolution of infrared LFs of different populations of infrared galaxies will be revealed in the future using the \textit{JWST} data, once its data become available in more filters.

\section{Conclusions}\label{S:conc}
With the deep MIR data from the \textit{JWST} ERO, we obtained the first source counts at 7.7 $\mu$m, 10.0 $\mu$m and 15.0 $\mu$m to characterise the source counts to the low flux limit of 0.4\,$\mu$Jy that has been unreachable with previous satellites. We measure 80\% completeness limits to be 0.40, 0.79, and 2.0\,$\mu$Jy in  F770W, F1000W and F1500W filters, respectively.
Our results agree well with models based on previous infrared satellite data (e.g., \textit{Spitzer} and \textit{AKARI}), confirming our understanding of the faint infrared population of galaxies. 

\section*{Acknowledgements}
The authors thank the anonymous referee for many insightful comments, which improved the paper much.
TG acknowledge the support of the National Science and Technology Council of Taiwan through grants 108-2628-M-007-004-MY3 and 110-2112-M-005-013-MY3. 
TH acknowledges the support of the National Science and Technology Council of Taiwan through grants 110-2112-M-005-013-MY3, 110-2112-M-007-034-, and 111-2123-M-001-008-.
This work is based on observations made with the NASA/ESA/CSA James Webb Space Telescope. The data were obtained from the Mikulski Archive for Space Telescopes at the Space Telescope Science Institute, which is operated by the Association of Universities for Research in Astronomy, Inc., under NASA contract NAS 5-03127 for \textit{JWST}. These observations are associated with program ERO.
The authors are grateful to Jamie Chang at the Department of Physics, NTHU for his dedication to the production of Figure \ref{fig:1}. His beautiful figures put the finishing touches to the paper.

\section*{Data Availability}
Early Release Observations obtained by \textit{JWST} MIRI are publicly available at \url{https://www.stsci.edu/jwst/science-execution/approved-programs/webb-first-image-observations}.
Other data underlying this article will be shared upon reasonable request to the corresponding author.



\bibliographystyle{mnras}
\bibliography{jwstpaper} 




%



\bsp	
\label{lastpage}
\end{document}